\documentclass{article}  
\usepackage{hip-artc}
\volnumber{} \issuenumber{} \edyear{2004}                                %|
\frompage{000} \topage{000}                                              %| 
\recrevdate{27 February 2004}                                                 %| 
\def\rightmark{What is the temperature \ldots} 
\def\leftmark{T.S.Bir\'o, G.Purcsel and B.M\"uller}
 
\title{{\bf What is the temperature in heavy ion collisions?}} 
\authors{
{Tam\'as S. Bir\'o$^1$, G\'abor Purcsel$^1$ and Berndt M\"uller$^{2}$%
\index{One, A.}
\index{Two, A.}
}\\[2.812mm]
{\normalsize
\hspace*{-8pt}$^1$ KFKI Res. Inst. Part. Nucl. Phys.\\ 
H-1525 Budapest, Pf.49, Hungary\\[0.2ex] 
\hspace*{-8pt}$^2$ Physics Dept. Duke University\\ 
Durham, NC-27708, USA
}}
 
\abstract{
 We consider the Tsallis distribution as a source of the apparent slope of 
 one-particle spectra in heavy-ion collisions and investigate the equation
 of state of this special kind of quark matter in the framework of
 non-extensive thermodynamics. We relate the energy per particle to the
 power-law tail of spectra at a given temperature.
}

 
\makeindex
\begin{document}
 
\maketitle
\setcounter{page}{1}
%%%%%%%%%%%%%%%%%%%%%%%%%%%%%%%% MAIN TEXT %%%%%%%%%%%%%

%\input{TS.tex}

%%%%%%%%%%%% ~/tex/notes/RHIC-School/TS.tex %%%%%%%%%%%%%
%
%	"thermal" model with Tsallis distribution
%
%	T.S.Biro	20.11.03
%	23.11.03	% re-written for q=1+1/c case, high-energy suppressed
%	27.11.03	% corrections, diff. type Tsallis approaches together
%	29.11.03	% thermodynamical relations, E/N expressions
%	25.02.04	% R!HIC-School 2003 talk, written version for APH
%	27.02.04	% include in template.tex for APH Heavy Ion Physics
%
%%%%%%%%%%%%%%%%%%%%%%%%%%%%%%%%%%%%%%%%

%\documentclass[12pt]{article}

%\usepackage{graphics}
%\usepackage{graphicx}
%\usepackage[dvips,usenames]{color}

\newcommand{\vs}{\vspace{1mm}}

\newcommand{\be}{\begin{equation}}
\newcommand{\ee}{\end{equation}}
\newcommand{\ba}{\begin{eqnarray}}
\newcommand{\ea}{\end{eqnarray}}
\newcommand{\NL}{\nonumber \\}

\newcommand{\PP}{\frac{d^3k}{(2\pi)^3} }

\parindent0mm
\fboxsep3mm
\fboxrule1mm

%\begin{document}

%\centerline{{\LARGE \bf What is the temperature in heavy ion collisions? }}
%\vs

%\centerline{{\large \sc Tam\'as S. Bir\'o, G\'abor Purcsel and Berndt M\"uller}}
%\vs
%\centerline{{\bf Version 25.Feb.2004.}}

%\vs

%\vs
%{\Large \bf Why temperature}

\section{Introduction}

\subsection{Why temperature}

\vs
There is a longstanding discussion in heavy-ion physics whether the concept
of temperature and the assumption of thermal equilibrium can be applied
for the description of the fireball formed before hadronisation in high-energy
reactions. We address here this question from the viewpoint of
very rough bulk properties relating the inverse slope of one-particle
transverse momentum spectra and the energy per particle.
Hadron mixture fits, the so called ''thermal model'' \cite{THMOD}
assumes a constant
temperature all over the fireball and obtains the energy per particle
using integrals with thermal distributions: basically the Gibbs distribution,
which leads to Fermi and Bose distributions in the grandcanonical treatment.
Although it has been shown that due to the finiteness of the available
phase space volume a canonical suppression may occur \cite{CANSUP}, the existence
of temperature was not questioned in this framework.

\vs
Study of the one-particle spectra gives a reason to believe in the temperature
as a universal parameter of the Gibbs distribution, $\exp(-E/T)$. From AGS
through SPS to RHIC experiments the transverse momentum spectra of
several particles, first of all pions, kaons and protons, show an exponential
fall. This occurs in its purest form in the $p_t=1-4$ GeV range.
At lower momenta deviations can be seen, traditionally attributed to
the presence of a global transverse flow. We do not intend to discuss this
part of the spectra here. At higher momenta in the RHIC experiment, however,
deviations occur too. Instead of the exponential curves a power-law
tail fits the experimental hadron data in the $p_t=4-12$ GeV range.

%\vs
%{\Large \bf Why no temperature}
\subsection{Why no temperature}

\vs
The power law tail in the pion spectrum contradicts to the assumption
of Gibbs distribution. Although one may consider the possibility that
only a minor part of the particles comes from a non-equilibrated source,
theoretically this doubt may be extended for the whole process.
Dynamical calculations indicate that the time may not be enough to
thermalize the whole spectrum before the break up \cite{DYNAMIC}.

\vs
In this paper we would like to test a model of hadronic and quark matter,
which considers a non-exponential one-particle distribution.
Such distributions with power-law tail belong to a class of anomalous
distributions, not satisfying the central limit theorem of statistics.
In particular the canonical Tsallis distribution, $(1+E/b)^{-c}$,
which has been derived by an attempt to generalize the traditional
thermodynamics  \cite{TSALLIS}, extrapolates between the expontial
curve for $E \ll b$ and the power-law tail for $E \gg b$. The apparent
temperature of the spectra is given by $T=b/c$. The Tsallis index
is related to the power as $q=1-1/c$ or $q=1+1/c$ depending
on the treatment of energy and particle number.

%\vs
%{\Large \bf Long tailed distributions}
\subsection{Long tailed distributions}

\vs
Before reviewing the main formulae of the non-extensive thermodynamics
and applying it for a massless quark-gluon plasma we remind to the
central limit theorem and show a possible exemption from it. This
exemption, the Lorentz distribution, has a Fourier transform corresponding
to exponential transverse mass spectra after regularization.

\vs
Considering $n$ independent random variables their properly scaled sum
is distributed more and more Gaussian; in the $n \rightarrow \infty$ limit
exactly Gaussian. The only requirement beyond independency is that
the fiducial variables must have a short range distribution.

%Giving an example uniform random variables in the interval $(-1,1)$, lead
%to a distribution of the sum $x=(\sqrt(3/n))\sum_i=1^n x_i$, which 
%approaches the Gaussian, $P_{\infty}(x)=(1/\sqrt(2\pi)) \exp(-x^2/2)$.
%The $j-th$ central moment of the distribution of the sum can be obtained
%in general as the $j-th$ derivative of the logarithm of the Fourier-transform
%of $P_n(x)$, $c_j(n)=(-i\partial/\partial k)^j \log \tilde{P}_n(k)$ at $k=0$.
%Considering the scaled sum, $x=a_n\sum_{i=1}^n x_i$, it is eeasy to derive
%that $c_j(n)=na_n^j c_j(1)$. For the choice $a_n=1/\sqrt{n}$, one arrive at
%$c_j(n) = n^{1-j/2} c_j(1)$ which goes to zero for $j > 2$ if $n$ is large.

\vs
In cases of unusual distributions, the central moments may diverge,
or just not vanish for large foldness $n$. 
%It may, however, be given another scaling
%property, $a_n$, which leads to a finite $c_j(n)$ in the $n\rightarrow\infty$
%limit. 
This is the case for the Lorentz-distribution, $w(x)=(T/\pi)/(1+T^2x^2)$.
Its Fourier transform is given by, $\tilde{w}(k)=\exp(-|k|/T)$, whose
logarithm cannot be differentiated at $k=0$. The distribution $P_n(x)$ of
$x=a_n\sum_{i=1}^n x_i$ has the Fourier transform $\tilde{P}_n(k)$ and
\be
 \log \tilde{P}_n(k) = - n |k| a_n/T.
\ee
The scaling $a_n=1/n$ leads in this case to a result independent of $n$
and hence to a limiting distribution, which happens to be the same Lorentzian.

\vs
Furthermore a physically supported regularization with a finite mass $m$
renders the central moments finite.
\be
 \tilde{w}(k) = \exp \left( \frac{m-\sqrt{k^2+m^2}}{T} \right)
\ee
leads to a distribution of the arithmetic sum $x=(1/n)\sum_{i=1}^n x_i$,
(phyiscally interpretable as a center of mass or energy), with a Fourier
transform,
\be
 \tilde{P}_n(k) = \tilde{w}^n(k/n) = 
  \exp \left( \frac{nm-\sqrt{k^2+(nm)^2}}{T} \right).
\ee
This may be the one-particle spectra of hadrons recombined from $n$
partons each with an almost vanishing mass $m \rightarrow 0$, but with
a finite total mass $M=nm$. A simple, but powerful picture of 
hadron formation from several partons, which leads to the experimentally
observed $\exp(M-M_t)$ transverse mass scaling spectra.

\vs
This cannot be, however, the whole story. Not only because of the power-law
tail of distributions, but also on theoretical grounds. Pure recombination,
as considered above, would need re-heating and expansion for not decreasing
the entropy during the spontaneous hadronization. Although no detailed
microdynamical description is yet known for this process, there are doubts,
whether enough expansion may take place for this entropy requirement.
Therefore it is of interest to circumline possible mechanisms which would
change the precursor quark matter distribution into another one of the
hadrons, closer to equilibrium and hence possesing higher entropy.

\vs
Such a case has been suggested in \cite{PLB2004}. The Tsallis distribution,
\be
 w_1(E) = \left( 1 + E/b \right)^{-c} = w_{TS}(E,T,1-1/c),
\ee
after $n$-fold recombination becomes another Tsallis distrinution,
\be
 w_n(E) = \left( 1 + E/nb \right)^{-nc} = w_{TS}(E,T,1-1/nc),
\ee
much closer to the Gibbs-limit $w_{\infty}(E)=w_{TS}(E,T,1)$.
It seems to be promising therefore to investigate the 
non-extensive thermodynamics of quark matter.

%\vs
%{\Large \bf Tsallis distribution and energy per particle}
\section{Tsallis distribution and energy per particle}

\vs
In this paper we concentrate on the question that how large the
energy per particle $E/N$ (simply related to the entropy per particle at
vanishing chemical potential) can become in a quark matter with
Tsallis distribution. Assuming a large enough volume for the fireball
and homogenecity, we take $E/N=e/n$, the corresponding ration of densities.
We shall compare this for the traditional relativistic Boltzmann gas
with massless particles, for the original bag-model description of
the quark-gluon plasma (QGP), and for a Tsallis-QGP with power-law 
tail distributed quasi-particles. We shall point out, that only the last
version is able to reproduce the high value of $E/N \approx 6T$ fitted
to RHIC experiments by the massive hadronic thermal model. This makes
the Tsallis-QGP to an interesting candidate for the pre-hadron quark
matter.

\vs
\subsection{Non-extensive thermodynamics}

\vs
In the following we breifly review basic formulae in the non-extensive
thermodynamics.
Tsallis has suggested an altered definition of entropy, which includes
the Boltzmannian case as a limit. Typically it leads to a power-law tailed
energy distribution instead of the exponential Gibbs-distribution, known
from equilibrium statistical physics. This seems to fit for
relativistic heavy ion collisions, where the observed transverse
momentum spectra of hadrons definitely show a power law tail.

\vs
The Tsallis entropy contains a parameter $q$, the familiar thermodynamics
is recovered in the limit $q=1$. It is given by

\be
 S_q  = \frac{1}{1-q}  \sum_i  (w_i^q - w_i)
\ee

with $w_i$ being the probability of the occurrence of the state $i$ in
the statistical sample of physical states. The parameter $q$ can be
related to a power characteristic to the high-energy part of these
probabilities by $q = 1+1/c$. The entropy can be expressed with the
help of $c$ as follows
\be
 S_q =  - \sum_i w_i c ( w_i^{1/c} - 1)
\label{TS-ENT}
\ee

\vs
The canonical probability is obtained by maximizing the entropy
(\ref{TS-ENT}) with a constraint on the energy and, if applies, on 
some charge conservation:
\be
 S_q - \beta ( E - \mu N) = {\rm max.} 
\ee
Instead the sum of the energies of the individual states,
$E' = \sum_i w_i E_i$, and corresponding charges $N' = \sum_i w_i Q_i$ 
in acertain version of Tsallis thermodynamics
one considers the q-weighted average quantities as thermodynamical
variables:
\be
 E = \frac{\sum_i w_i^qE_i}{\sum_i w_i^q}, \qquad \qquad
 N = \frac{\sum_i w_i^qQ_i}{\sum_i w_i^q}.
\ee
One assumes that these $E$ and $N$ are additive for subsystem and reservoir
parameterizing this way non-linearities (correlations) in energy and
particle number. (Tsallis brings only one argument in favor of this choice
against the naive one: the latter "turns out to be inadequate for
various reasons, including related to L\'evy-like superdiffusion, for which
a diverging second moment exists". No reference found to this, yet.)

\vs
The optimal canonical distribution can so be obtained as
\be
 w_i = \frac{1}{Z} \exp_{c}(-x_i),
\ee
with $x_i = (E_i-\mu Q_i)/T$.
Here we introduced the parameter $T$, which we refer to as "temperature"
in the following discussion. This is, however, not equal to
the inverse of the Lagrange multiplier $\beta$ \cite{COND}.

\vs
The basic thermodynamical relation can be expressed, as follows:
\be
 -\Omega = pV = T \ln_c Z = T S - (E - \mu N) (1-S/c)
\label{TS-TH1}
\ee
with $\ln_c(x)=c(1-x^{-1/c})$.
This relation defines the pressure, and
nicely resembles the thermodynamics in the usual form
with corrections in the order of $1/c$. 

\vs 
The pressure is a function of $T$ and $\mu$ only in the homogeneous limit. 
The derivatives of the grandcanonical thermodynamical potential are
\be
\frac{\partial}{\partial V} pV = p, \qquad \qquad
\frac{\partial}{\partial T} pV  = S, \qquad \qquad
\frac{\partial}{\partial \mu} pV = N(1-S/c), 
\label{TS-DER}
\ee
This helps to re-express the energy and the particle number in terms
of the pressure and its derivatives:
\be
 N =  V \frac{ \frac{\partial p}{\partial \mu} }{ 1 - \frac{V}{c}
   \frac{\partial p}{\partial T} }, \qquad \qquad
 E = V \frac{ \mu \frac{\partial p}{\partial \mu} 
  + T \frac{\partial p}{\partial T}  - p }{ 
               1 - \frac{V}{c} \frac{\partial p}{\partial T} }.
\label{TS-E}
\ee
It is interesting to note, that the energy per particle $E/N$ does not
contain the ${\cal O}(1/c)$ correction factor directly, it resembles
the usual expression
\be
 \frac{E}{N} =
  \frac{ \mu \frac{\partial p}{\partial \mu} 
  + T \frac{\partial p}{\partial T}  - p }{ \frac{\partial p}{\partial \mu} }.
\label{TS-EperN}
\ee

\vs
We note here, that using the $E'=\sum w_iE_i$ average energy for the canonical
constraint (and similarily $N'= \sum w_i Q_i$) leads right away to the
same distribution,
\be
 w_i = \frac{1}{Z} \left( 1 + x_i/c \right)^{-c},
\ee
but now $q = 1-1/c$ (the relation to the Tsallis index changes).
Also the thermodynamics, defining $pV$ by $T \ln_{c} Z$ contains finite
${\cal O}(1/c)$ modifications:
\be
 TS = pV ( 1 + S/c) + E - \mu N.
\ee
Now also the derivatives of the grandcanonical potential receive
corresponding corrections, but the expressions for $E$ and $N$
are formally the same, as in the other case. The relation of the entropy
to the pressure, however, changes,
\be
 S =  V \frac{ \frac{\partial p}{\partial T} }{ 1 - \frac{V}{c}
   \frac{\partial p}{\partial T} },
\label{TS-S}
\ee
as well as the definition of pressure now belongs to a $q$ index reflected
to $1$.

\vs
Summarizing, irrespective to which Tsallis-case we choose, we define the
pressure and the grandcanonical potential correspondingly, so that in
both cases
\be
p V = T c ( 1 - Z^{-1/c}).
\ee
The canonical distribution shows a power-law tail:
\be
  w_i = \frac{1}{Z}  \frac{1}{\left(1+x_i/c \right)^c}.
\label{TS-CAN}  
\ee
The partition sum $Z$ is obtained by requiring $\sum_i w_i = 1$.

%\vs
%{\bf Tsallis ideal gas, bosons, fermions}
\subsection{Tsallis ideal gas, bosons, fermions}

\vs
In the homogeneous case, very often encountered in simple physics models,
the partition function is considered as a product of contributions from
each particle state with a given momentum $\vec{k}$. The familiar logarithm
of such a partition function is then additive in the phase space.

\vs
In the case of the non-equilibrium Tsallis-distribution the entropy is,
however, no more additive; the above mechanism is no more trivial.
Assuming that the energy, volume and particle number are still additive, 
the Tsallis ideal gas the following pressure:
\be
  p(\mu,T) = \int \PP T  \ln_{c} Z,
\ee
with $Z$ being the one-particle Tsallis partition sum,
\be
  Z = \sum_n \exp_{c}(-nx),  \qquad \qquad  x = (E_k-\mu)/T.
\label{1PT-Z}
\ee
here $\exp_c(x)$ is the inverse function of $\ln_c(x)$.
This is the basis of the distinction between Fermi-Dirac and
Bose-Einstein statistics: in the former case only $n=0$ and $n=1$
contributes to the sum, while in the letter case all non-negative
$n$-s from zero to infinity.

\vs
The $1/c$-uncorrected derivatives of the pressure give,
\be
 n_0 = \int \PP n_c(\frac{E_k-\mu}{T}), \qquad
 e_0 = \int \PP E_k n_c(\frac{E_k-\mu}{T})
\ee
with
\be
 n_c(x) = - \frac{\partial}{\partial x} \ln_{c}Z(x).
\ee

\vs
Let us now analyze the one-particle distribution and the partition
sum. First we regard the Boltzmann limit, $\exp_{c}(-x) \ll 1$.
We get
\be
 n_c(x) \approx -\frac{d}{dx} \ln_{c}(1+\exp_{c}(-x))
 \approx -\frac{d}{dx} \exp_{c}(-x) = \left( 1 + \frac{x}{c} \right)^{-c-1}.
\ee
It is seen, that the Tsallis temperature, $T$ and the inverse slope of the
logarithmic one-particle spectra somewhat differ for finite $c$ in this
scenario:
\be
 (c+1) T_{{\rm slope}} = c T = b. 
\ee
The product quantity, $b$ is regarded as an energy (or transverse momentum)
scale which characterizes the onset of the power-law tail of the canonical Tsallis
distribution. The general shape is given by
\be
  n_c(E) = \left( 1 + E/b \right)^{-c-1}
\ee
for vanishing chemical potential $\mu=0$. In this case the particle
number density and energy density integrals can be calculated analytically,
resulting
\be
  n_0 = \frac{1}{2\pi^2} \frac{2b^3}{c(c-1)(c-2)}, \qquad \qquad
  e_0 = \frac{1}{2\pi^2} \frac{6b^4}{c(c-1)(c-2)(c-3)}
\ee
This suffices to express a relation between the inverse slope of the
one-particle energy spectrum at small energy $E \ll b$, 
the energy per particle and the Tsallis parameter q=1+1/c:
\be
  E/N = e/n = \frac{3q}{4-3q} \;  T_{{\rm slope}}.
\ee
An estimate can be given from the RHIC experiments, pointing out an
inverse slope of $175$ MeV (cleaned from transverse flow effects) and
$E/N = 1$ GeV. This agrees with $q=1.135$ in the Boltzmann ideal gas limit
of massless partons, a much more likely candidate for the pre-hadron
matter, then the usual QGP:  at the same $E/N$ value the massless Bose
distribution leads to $T = 325$ MeV temperature and slope.

%%%%%%%%%%%%%%%% FIG.1 %%%%%%%%%%%%%%%

\begin{figure}[htb]
\vspace*{-11mm}
 \insertplot{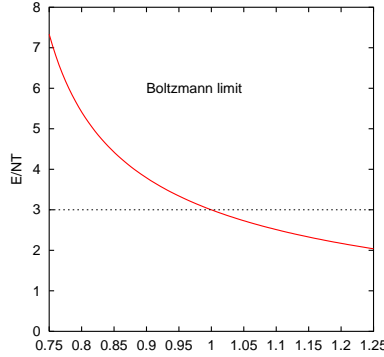}
\vspace*{-21mm}
\caption[]{
 Energy per particle divided by the temperature ($E/NT$) as a function
 of the Tsallis index.
}
\label{FIG1}
\end{figure}
%%%%%%%%%%%%%%%%%%%%%%%%%%%%%%%%%%%%%

%\vs
%{\bf \large Finite chemical potential}
\subsection{Finite chemical potential}

\vs
We are interested in the constant $E/N$ curve on the $T-\mu$ plane for
a mixture of massless bosons and fermions, which are Tsallis-distributed.
Denoting $(E-qmu)/T$ generally by $x$ we deal with the following type of
integral experessions:
\be
 n_0 = \int dk \; k^2 n_c(x) = \int_{-\mu/T}^{\infty} \! dx \;
 (xT+\mu)^2 n_c(x).
\ee
Similarily for the quantity $e_0$, just a higher power of $(xT+\mu)$
occurs under the integral sign. The one-particle distributions
$n_c(x)$ are for fermions or for bosons respectively.

\vs
Denoting the general integrals by
$I_n(a) = \int_{-a}^{\infty} dx \; x^n n_c(x)$,
the result consists of different power contributions. For massless fermions
and antifermions together we get
\be
n_0 = T^3 S_2(a) + 2T^2\mu D_1(a) + T\mu^2S_0(a),
\ee
and
\be
 e_0 = T^4 S_3(a) + 3T^3\mu D_2(a) + + 3T^2\mu^2S_1(a)  + T\mu^3D_0(a).
\ee
with $S_n(a)=I_n(a)+I_n(-a)$ and $D_n(a)=I_n(a)-I_n(-a)$, $a=\mu/T$. 
For bosons in chemical equilibrium one takes $\mu=0$, so only
the leading integrals remain, of course, in this case with Bose-Tsallis
$n_c(x)$ functions. The energy per particle for a mixture, like quark matter,
becomes
\be
 \frac{E}{N} = 
 \frac{d_B e_{0B}(T) + d_F e_{0F}(T,\mu) + d_F e_{0F}(T,-\mu)}{d_B n_{0B}(T) + d_F n_{0F}(T,\mu) + d_F n_{0F}(T,-\mu)}.
\ee
For $c \rightarrow \infty$ one is back with the familiar Fermi-Dirac and
Bose-Einstein distribution functions, and both $e_0$ and $n_0$ are finite
order polynomials of $T$ and $\mu$. The ratio is a growing function of both,
so a constant $E/N$ reflects in a convex curve in the $T-\mu$ plane.
For large enough $c$ the Tsallis version has to behave similarily, but this
question we can only decide numerically.

%%%%%%%%%%%%%%%%%%%%%%% FIGURES 2 and 3  %%%%%%%%%%%%%%%%%%%%%

\begin{figure}[htb]
\vspace*{-11mm}
 \insertplot{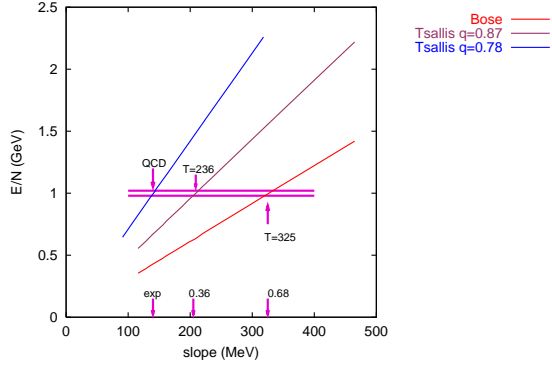}
\vspace*{-21mm}
\caption[]{
 Energy per particle for massless Boson gas with Tsallis and Gibbs 
 distributions as function of the spectral slope.
}
\label{FIG2}
\end{figure}

\begin{figure}[htb]
\vspace*{-11mm}
 \insertplot{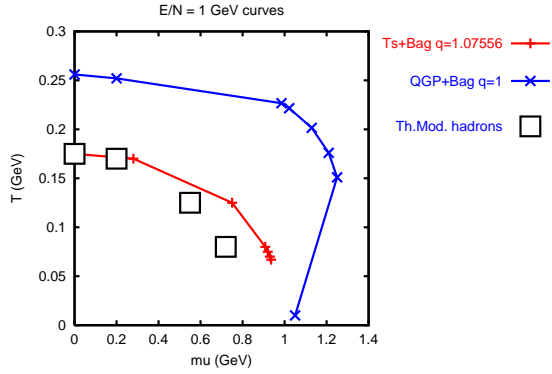}
\vspace*{-21mm}
\caption[]{
 The $E/N = 1$ GeV line on the temperature -- chemical potential plane
 for the Gibbs and for the Tsallis distribution.
}
\label{FIG3}
\end{figure}

%%%%%%%%%%%%%%%%%%%%%%%%%%%%%%%%%%%%%%%%%%%%%%%%%%%%%%%%%%%%%%%%%

%\vs
%{\bf \Large Comparison of Quark Matter Models}
\section{Comparison of Quark Matter Models}

\vs
In conclusion we have compared three models for quark matter.
The simplest, the massless Boltzmann gas, as it is long known,
gives $E/N = e/n = 3T$. 
The equation of state is given by,
\be
 p = \frac{1}{3} \sigma T^4,  \qquad e = \sigma T^4, \qquad p = nT.
\ee

\vs
Supporting the Stefan-Boltzmann gas with a bag constant, $B$, 
we arrive at the traditional
bag-model QGP. Now the situation is a little bit more complex.
The equation of state is modified,
\be
 p = \frac{1}{3} \sigma T^4 - B,  \qquad e = \sigma T^4 + B, \qquad p = nT - B.
\ee
This kind of matter has a stability limit at $p=0$, so the bag constant
is related to a critical temperature $B=\sigma T_{cr}^4/3$. The energy
per particle becomes at this limit $E/N = e_{cr}/n_{cr} = 4T$.

\vs
Finally in a Tsallis-distributed quark matter the Stefan-Boltzmann constant
$\sigma$ depends on the power $c$ (or Tsallis index  $q=1-1/c$). The
equation of state becomes
\be
 p = \frac{1}{3} \sigma(c) T^4,  \qquad e = \sigma(c) T^4, 
      \qquad n = \lambda(c)T^3.
\ee
The energy per particle in the Boltzmann limit is given by
\be
 E/N = \frac{c}{c-4} 3T,
\ee
which is about $6T$ for $c=8$. In a bag constant supported Tsallis-QGP
version one obtains, $E/N = 4cT/(c-4)$ and a power of $c=12$.

%\end{document}

%%%%%%%%%%%%%%%%%%%%%%%% ACKNOWLEDGEMENT %%%%%%%%%%%%%%%%%%%%
 
\section*{Acknowledgement(s)}
This work has been supported by the Hungarian National Research Fund,
OTKA (T034269).

%%%%%%%%%%%%%%%%%%%%%%%%%%%%%%% REFERENCES %%%%%%%%%%%%%%%

\vfill\eject
\end{document}